\documentclass[showpacs,aps,floats,epsf,psfig,prc,twocolumn,graphics]{revtex4}
\usepackage{graphicx}
\renewcommand{\thetable}{\Roman{table}}

\begin{document}

\title{Solar Neutrinos from CNO Electron Capture}

\author{L.~C.~Stonehill, J.~A.~Formaggio, and R.~G.~H.~Robertson} 

\affiliation{ 
Center for Experimental Nuclear Physics and Astrophysics, and Department of
Physics\\
University of Washington, Seattle, WA 98195 \\
}
\date{\today}         

\begin{abstract}
The neutrino flux from the sun is predicted to have a CNO-cycle contribution as
well as the known $pp$-chain component. Previously, only the fluxes from $\beta
^{´+}´$ decays of $^{´13}´\rm N$, $^{´15}´\rm O$, and $^{´17}´\rm F$ have been
calculated in detail.  Another neutrino component that has not been widely considered is
electron capture on these nuclei. We calculate the number of interactions in
several solar neutrino detectors due to neutrinos from electron capture on
$^{´13}´\rm N$, $^{´15}´\rm O$, and $^{´17}´\rm F$, within the context of the
Standard Solar Model. We also discuss possible non-standard models where the
CNO flux is increased.
\end{abstract}
\pacs{23.40.-s, 26.65.+t, 14.60.Lm}
\maketitle
\section{Introduction}

Experimental data gathered from both radiochemical
\cite{Homestake,Sage,Gallex,GNO} and real-time solar neutrino experiments
\cite{SNO,SK} not only have revealed the phenomena of neutrino oscillations,
but also have established the predominant mechanism for solar fuel burning
\cite{bib:BP}.  The driving component for nuclear burning in the sun is the
$pp$ fusion chain.  However, it is predicted that a portion of the solar
neutrino flux also comes from the CNO cycle \cite{Bethe}.  The CNO reaction
products that have been shown to produce significant neutrino fluxes include
$\beta ^{´+}´$ decays of $^{´13}´\rm N$, $^{´15}´\rm O$, and $^{´17}´\rm F$. 
However, an additional source of neutrinos not previously evaluated in detail is electron capture on $^{´13}´\rm N$, $^{´15}´\rm O$, and $^{´17}´\rm F$.  Electron capture produces a mono-energetic line spectrum with energy 1.022 MeV above the endpoint of the $\beta ^{´+}´$ continuum.  Bahcall \cite{bib:line} has considered electron capture from free electrons in the solar plasma, but not bound state electrons.

The increased sensitivity and precision of current and future solar neutrino
experiments make it difficult to ignore contributions from these reactions. 
Moreover, the existence of a line spectrum presents an opportunity to make
precision measurements of CNO fluxes.  Existing solar neutrino experiments are
sensitive to these neutrinos; in particular, the Sudbury Neutrino Observatory
(SNO) is sensitive to the higher-energy CNO neutrinos produced from electron
capture but not to the $\beta ^{´+}´$ continuum.  This contribution must be
estimated in order to make a correct assessment of the $^8$B flux -- not only its
magnitude, but also its spectral shape in the low-energy regime where matter
effects are expected.  In this paper, we calculate the predicted contribution
to present and future solar neutrino experiments from CNO electron capture
neutrinos. In addition, we also discuss cases of non-standard solar models in
which the CNO flux is increased.

\section{Electron Capture Fluxes}

The electron capture processes that occur in the CNO cycle involve the
following reactions:
\begin{eqnarray}
^{´13}´{\rm N} + e^{´-}´ \rightarrow {^{´13}´{\rm C}} + \nu_e \\
^{´15}´{\rm O} + e^{´-}´ \rightarrow {^{´15}´{\rm N}} + \nu_e \\
^{´17}´{\rm F} + e^{´-}´ \rightarrow {^{´17}´{\rm O}} + \nu_e 
\end{eqnarray}

If the electron capture process is dominated by bound electrons, then it is
possible to relate the electron capture flux directly to the $\beta ^{´+}´$
decay flux \cite{bib:NDT}. At solar temperatures and densities, however, one
must take into account the contribution from both bound and continuum
electrons.  The ratio between electron capture rates in the sun and laboratory
measurements is given by \cite{bib:ikc}:
\begin{equation}
R \equiv \frac{\lambda_{\rm sun}}{\lambda_{\rm lab}} = n_e ~
\frac{~|\psi(0)_{\rm sun}|^2}{2 |\psi(0)_{\rm lab}|^2},
\end{equation}
where $n_e$ is the electron density in the sun and the atomic wave functions
$\psi$ are given by:
\begin{equation}
|\psi(0)_{\rm lab}|^2 = \frac{1}{\pi} Z^3 \kappa(Z)
\end{equation}
\begin{equation}
|\psi(0)_{\rm sun}|^2 = \exp{(-\frac{Z \beta}{R_D})}(\omega_c + \omega_b).
\end{equation}

Here $Z$ is the charge, $\kappa(Z)$ is the correction term applied to the pure
Coulomb field of $4Z^3\alpha^3$, as tabulated in \cite{bib:isotope}, $\beta
\equiv \frac{1}{kT}$ is expressed in units of $\hbar = e = m_e =1$ \cite{foot},
and $T$ is the solar temperature. The factors $\omega_c$ and $\omega_b$ are
continuum and bound state electron density ratios at the nucleus for
Coulomb-distorted waves relative to plane waves.  Also included is a weak solar
plasma screening correction which depends on the Debye radius ($R_D$)
\cite{bib:Salpeter}.  The continuum and bound state electron density ratios are
given by \cite{bib:bahcall97}:
\begin{equation}
\label{eqn:wc}
\omega_c = <\frac{2\pi \eta}{1-e^{-2\pi\eta}}>
\end{equation}
\begin{equation}
\label{eqn:wb}
\omega_b = \pi^{\frac{1}{2}} (2Z^2 \beta )^{\frac{3}{2}} \sum \frac{1}{n^3}
\exp{(\frac{Z^2 \beta}{2n^2})},
\end{equation} 
where $\eta=Z/v$ is the inverse velocity averaged over the electron
Maxwell-Boltzmann distribution.  

The electron density ratios are evaluated at both a fixed point in the solar
core ($R_0$), and integrated over the entire solar volume ($R_{\infty}$).  The
fixed point used is 0.057 of the solar radius, where the $^{´13}´{\rm
N}$, $^{´15}´{\rm O}$, and $^{´17}´{\rm F}$ fluxes peak. At this location, the
temperature is $1.48 \times 10^7$~K, the Debye radius is 0.45, and the density
is $5.32 \times 10^{25}$ atoms/cm$^3$ \cite{bib:BP}. The effect of the full
integration on the fluxes is small for the nuclei of interest ($\sim$ 3\% for
$^{´13}´{\rm N}$ and less than 1\% for $^{´15}´{\rm O}$ and $^{´17}´{\rm F}$).
The total correction due to continuum electron capture is shown in
Table~\ref{tab:kshell}.  The relative K-shell/L-shell occupancy for $^{´13}´\rm
N$, $^{´15}´\rm O$, and $^{´17}´\rm F$ are all greater than 90\%
\cite{bib:isotope}.  Capture of both K- and L-shell electrons has been included
here.  For evaluation of the electron capture rate with accuracy of a few percent
the radiative corrections should be included (see, for example, \cite{Kurylov}).  

\setlength{\tabcolsep}{0.10in}
\begin{table}[h]
\caption{The fraction of bound state electrons in the solar core, the atomic
wave function at the nucleus in the sun, and the total correction to the
electron capture rate. Both fixed point ($R_0$) and volume-integrated
($R_{\infty}$) ratios are shown. $^{´7}´{\rm Be}$ is shown for comparison.}
\label{tab:kshell}
\begin{center}
\begin{tabular}{ccccc}
\hline
\hline
Element   & $\omega_b / (\omega_c+\omega_b)$ & $|\psi(0)_{\rm sun}|^2$ & $R_0$
& $R_{\infty}$\\
\hline
$^{´7}´{\rm Be}$ & 0.302   &   3.76   &  0.858  &  0.804 \\
$^{´13}´{\rm N}$ & 0.662   &  11.08   &  0.419  &  0.403 \\
$^{´15}´{\rm O}$ & 0.749   &  16.14   &  0.400  &  0.398 \\
$^{´17}´{\rm F}$ & 0.818   &  23.75   &  0.406  &  0.405 \\
\hline
\hline
\end{tabular}
\end{center}
\end{table}

\begin{figure}[htb]
\begin{center}
\includegraphics[width = \columnwidth,keepaspectratio=true]{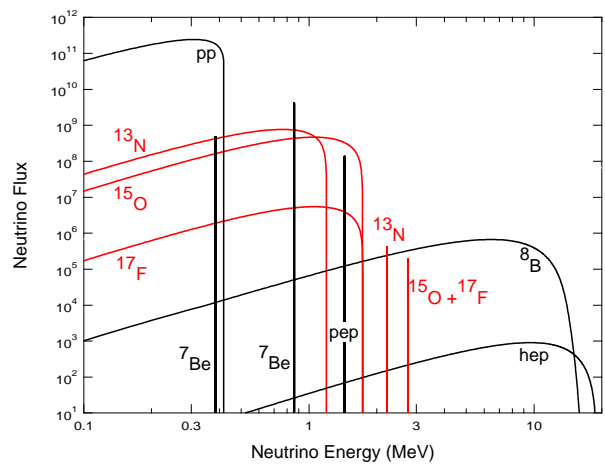}
\caption{Solar neutrino flux at 1 A.U., including electron capture in the CNO
cycle.  The $pp$ chain is shown in black and the CNO cycle is shown in red. 
Line fluxes are in cm$^{-2}$ s$^{-1}$ and spectral fluxes are in cm$^{-2}$
s$^{-1}$ MeV$^{-1}$ The $pp$ chain and the CNO $\beta ^{´+}´$ decay fluxes are
from \cite{bib:Bahcallwebsite}.}
\label{fig:ec}
\end{center}
\end{figure}

Table~\renewcommand{\thetable}{\Roman{table}}\ref{tab:fluxes} shows the
expected total rate of neutrinos from K-shell and continuum electron capture
processes, assuming the solar burning cycle is dominated by $pp$ fusion.  The
major contribution to the uncertainties on the electron capture fluxes comes
from the uncertainties on the standard solar model (SSM) $\beta ^{´+}´$ decay
fluxes \cite{bib:BP}. The neutrino flux from these sources is of the same order
as the $^{´8}´\rm B$ flux, though at lower neutrino energies.  The solar
neutrino spectrum, including the CNO electron capture neutrino lines, is shown
in Fig.~\ref{fig:ec}.  There is in addition an electron-capture branch for $^8$B decay \cite{bib:Villante}, but its total flux is 1.3 cm$^{-2}$ s$^{-1}$, too small to appear on the graph.

\setlength{\tabcolsep}{0.055in}
\begin{table}[htb]
\caption{Neutrino fluxes from CNO electron capture. The final electron capture
flux takes into account the correction for capture of continuum electrons
($R_\infty$).  The CNO cycle is assumed to be at the level dictated by the
SSM.} 
\label{tab:fluxes}
\begin{center}
\begin{tabular}{cccc}
\hline
\hline
 & SSM $\beta ^{´+}´$ decay flux & (EC/$\beta ^{´+}´$ decay)$_{\rm lab}$ & EC
flux\\ 
 & ($\rm cm^{´-2} \, s^{´-1}´$) &  & ($\rm cm^{´-2} \, s^{´-1}´$)\\
\hline
$^{´13}´\rm N$ & $5.48 \times 10^{´8}´ \, (^{+0.21 \%}_{-0.17 \%})$ & $1.96
\times 10^{´-3}´$ & $4.33 \times 10^{´5}´ $\\ [0.05in]
$^{´15}´\rm O$ & $4.80 \times 10^{´8}´ \, (^{+0.25 \%}_{-0.19 \%})$ & $9.94
\times 10^{´-4}´$ & $1.90 \times 10^{´5}´ $\\ [0.05in]
$^{´17}´\rm F$ & $5.63 \times 10^{´6}´ \, (^{+0.25 \%}_{-0.25 \%})$ & $1.45
\times 10^{´-3}´$ & $3.32 \times 10^{´3}´ $\\
\hline
\hline
\end{tabular}
\end{center}
\end{table}

To determine the observed rate at a given experiment, we consider charged
current (CC), neutral current (NC), and elastic scattering (ES) interactions on
a variety of targets.  The cross-sections used for $^2$H, $^7$Li, $^{37}$Cl and
$^{71}$Ga are taken from \cite{BCK,bib:SNOxsec,LIxsec,bib:CLGAxsec}.  A precise accounting of radiative corrections has not yet been applied to all of these processes.   In the case of deuterium targets, the value of L$_{1,A}$ \cite{BCK} was set to 4.0 and radiative corrections are included \cite{Kurylov}.
For the elastic scattering cross-section, the following relation was used:

\begin{equation}
\frac{d\sigma(\nu_e e^-)}{dT_e} = \frac{G_F^2 s}{\pi}\{g_L^2 + g_R^2
(1-\frac{T_e}{E_\nu})^2- g_L g_R \frac{m_e T_e}{E_\nu^2}\}
\end{equation}

\noindent where $G_F$ is the Fermi constant, $s$ is the center-of-mass energy,
$g_{L(R)}$ are the left (right) handed couplings for the weak current, $T_e$ is
the electron kinetic energy, and $E_\nu$ is the neutrino energy. Uncertainties
on electron, $^2$H and $^{7}$Li targets are well understood at the level of 1\%
\cite{SNOsalt}. Uncertainties in the CC cross-sections for $^{37}$Cl are
dominated by transitions to forbidden states, which at these energies are
1-2\%.    For CC interactions on $^{71}$Ga, allowed transitions to excited
states play a significant role and the uncertainties are expected to be larger
at these energies. The expected neutrino rates for various targets are
presented in Table~\renewcommand{\thetable}{\Roman{table}}\ref{tab:XSec1}. The
relatively large rates suggest that a $^7\rm Li$-based detector might be a
viable next-generation solar neutrino experiment.  For example, a water
Cherenkov detector with dissolved $^7\rm Li$, such as suggested in
\cite{Haxton}, might be a workable design.  

\setlength{\tabcolsep}{0.13in}
\begin{table*}[t]
\caption{Neutrino interaction rates with various detector materials, assuming
no neutrino oscillations. Rates are given in units of SNU's (1 SNU $\equiv
10^{-36}$ interactions/atom/s), except for ES, which is given in $10^{-36}$
interactions/electrons/s.}
\label{tab:XSec1}
\begin{center}
\begin{tabular}{cccccccc}
\hline
\hline
 & Energy (MeV)&  ES  &   $^2$H NC    &   $^2$H CC   &  $^7\rm Li$       & 
$^{´37}´\rm Cl$    &  $^{71}\rm Ga$       \\
\hline
$^{13}\rm N$ & 2.220 & $7.98 \times 10^{-3}$ & 0 & $3.63 \times 10^{-3}$ &
$8.79 \times 10^{-2}$ & $2.11 \times 10^{-3}$ & $2.15 \times 10^{-2}$  \\
$^{15}\rm O$ & 2.754 & $4.46 \times 10^{-3}$ & $2.26 \times 10^{-4}$ & $5.79
\times 10^{-3}$  & $6.65 \times 10^{-2}$ & $1.60 \times 10^{-3}$ & $1.54 \times
10^{-2}$ \\
$^{17}\rm F$ & 2.761 & $7.80 \times 10^{-5}$ & $4.08 \times 10^{-6}$ &  $1.02
\times 10^{-4}$ & $1.17 \times 10^{-3}$ & $2.80 \times 10^{-5}$ &  $2.70 \times
10^{-4}$ \\
\hline
\hline
\end{tabular}
\end{center}
\end{table*}

Of particular interest is whether the CNO electron capture flux constitutes a
serious background for current neutrino experiments.  For the Sudbury Neutrino
Observatory (SNO), these NC rates correspond to about 0.4 $^{´15}´\rm O$
neutrino NC event per year and about 0.01 $^{´17}´\rm F$ events per year.  The
latter is negligible, but the $^{´15}´\rm O$ contributes a small
model-dependent background to the $^{´8}´\rm B$ measurement.  The CC
interactions are below the SNO analysis threshold, so they do not contribute
significantly to SNO results.  Below the 5.5 MeV analysis threshold in the
recent SNO publication \cite{SNOsalt} there were about 13 events expected from
this source.  The ES interactions could be detected in a liquid scintillator
experiment like KamLAND \cite{bib:KamLAND} or BOREXINO \cite{bib:Borexino}. 
For example, in BOREXINO the electron capture neutrino rates would be about
0.1\% of the expected SSM signal. The expected rates for $^{71}\rm Ga$ and
$^{´37}´\rm Cl$ have also been calculated and are shown in
Table~\ref{tab:rates}.

\section{Alternative Solar Models}

This calculation has assumed that the CNO-cycle contribution to the solar
luminosity is 1.5\%, as predicted by the standard solar model \cite{bib:BP}. 
That model is well established theoretically, and fits well with
helioseismology data and the total $^{´8}´\rm B$ solar neutrino flux measured
by SNO.  It is possible, however, to envision other solar models in which the
CNO cycle is increased relative to the $pp$ chain, while still fitting with
available experimental data.  For example, the authors of \cite{bib:BFK}
suggested a model in which 99.95\% of the solar energy comes from the CNO cycle
while still agreeing with solar luminosity and the neutrino measurements to
that date.  In that model the $^{´15}´\rm O$ $\beta ^{´+}´$ decay flux is $3.41
\times 10^{´10}´ \, \rm cm^{-2} \, s^{´-1}´$, a 70-fold increase over the SSM
flux, which would raise the predicted $^{´15}´\rm O$ electron capture neutrino
NC rate in SNO to 30 $\rm yr^{´-1}´$.  This increase in the CNO flux does not
come at the expense of $^{´8}´\rm B$ flux, as the $^{´8}´\rm B$ flux in the
model is $8.64 \times 10^{´6}´ \, \rm cm^{-2} \, s^{´-1}´$, even higher than
the flux measured by SNO.  The model was not proposed as a realistic solar
model, rather it was an illustration of the possible level to which the CNO
cycle could be raised in the sun.  

Recent experimental results \cite{Homestake,Sage,Gallex,GNO,SNO,SK,bib:KamLAND}
constrain the fraction of energy that the sun produces via the CNO cycle to
less than 7.3\% at 3$\sigma$ \cite{bib:BGP}.  CNO electron capture neutrino
interaction rates in various neutrino detectors are shown in
Table\renewcommand{\thetable}{\Roman{table}}~\ref{tab:rates}, in the context of
the SSM as well as the 7.3\% upper limit model and the 99.95\% model.  Future
low-energy, high-resolution neutrino experiments can take advantage of the
electron capture channels to explicitly set more stringent limits on the
fraction of CNO neutrinos.

\section{Conclusion}

The neutrino flux from electron capture in the solar CNO-cycle
has been calculated. The rate of such neutrinos on current detectors is
expected to be small, though the process does introduce a model-dependent
background to the SNO measurement of the total $^{´8}´\rm B$ flux, at the level
of about one event per year.  However, the model-dependence is small, since the
fractional contribution of the CNO cycle to the solar luminosity is limited
experimentally to 7.3\%, only about a factor of five above the SSM fraction. 
Future experiments can take advantage of the mono-energetic nature of the
neutrinos from electron capture to make a precision measurement of the fraction
of the solar luminosity due to the CNO cycle.

\setlength{\tabcolsep}{0.13in}
\begin{table}[h]
\caption{CNO electron capture neutrino interaction rates in various detectors. 
Rates are presented for the SSM CNO fraction, the upper limit to the CNO
fraction that comes from solar neutrino data, and a toy model where almost all
of the solar luminosity is due to the CNO cycle.  Rates are given as a fraction
of the observed rate, except for BOREXINO, which is given as a fraction of the
expected rate.}
\label{tab:rates}
\begin{center}
\begin{tabular}{cccc}
\hline
\hline
 & SSM & 7.3\% & 99.95\%\\
\hline
SNO NC (Salt Phase) & 0.01\% & 0.05\% & 0.6\% \\
BOREXINO  & 0.1\%  & 0.3\% &  4.3\% \\ 
$^{37}$Cl & 0.2\%  & 0.7\% & 9.7\% \\
$^{71}$Ga & 0.1\%  & 0.2\% &  3.3\% \\ 
\hline
\hline
\end{tabular}
\end{center}
\end{table}

\section{Acknowledgments}
The authors would like to thank M.~K.~Bacrania for his assistance in preparing
Figure~\ref{fig:ec}. This work was supported by the U.~S.~Department of Energy
under Grant No. DE-FG06-90ER40537.


\begin{thebibliography}{99}
\bibitem{foot} With this convention, the unit of energy is the Hartree energy
($m_e c^2 \alpha ^2$) and the unit of length is the Bohr radius ($a_0$).
\end{thebibliography}
\end{document}